\documentclass[%
 reprint,
 amsmath,amssymb,
 aps,
]{revtex4-2}

\usepackage{graphicx}
\usepackage{dcolumn}
\usepackage{bm}
\usepackage{setspace}
\usepackage{ulem}
\usepackage{xcolor}

\usepackage{cleveref}

\begin{document}
\preprint{APS/123-QED}

\title{Boltzmann Bridges}

\author{Jordan Scharnhorst${}^{ab}$}
\email[]{Corresponding author: jscharnh@ucsc.edu}

\author{David Wolpert${}^a$}

\author{Carlo Rovelli${}^{acdef}$} 


\affiliation{\vspace{.25cm}${}^a$ Santa Fe Institute, 1399 Hyde Park Road
Santa Fe, NM 87501, USA}
\affiliation{${}^b$ Department of Physics, University of California, Santa Cruz, CA 95060, USA}
\affiliation{${}^c$ Department of Philosophy, University of Western Ontario, London, ON N6A 3K7, Canada}
\affiliation{${}^d$ The Rotman Institute of Philosophy, 1151 Richmond St.~N London  N6A5B7, Canada}
\affiliation{${}^e$ Perimeter Institute, 31 Caroline Street N, Waterloo ON, N2L2Y5, Canada} 
\affiliation{${}^f$ AMU Universit\'e, Universit\'e de Toulon, CNRS, CPT, F-13288 Marseille, EU}

\date{\today}

\begin{abstract}
\noindent It is often stated that the second law of thermodynamics follows from the condition that at some given time in the past the entropy was lower than it is now. Formally, this condition is the statement that $E[S(t)|S(t_0)]$, the expected entropy of the universe at the current time $t$ conditioned on its value $S(t_0)$ at a time $t_0$ in the past, is an increasing function of $t  $. We point out that in general this is incorrect. The epistemic axioms underlying probability theory say that we should condition expectations on all that we know, and on nothing that we do not know. Arguably, we know the value of the universe's entropy at the present time $t$ at least as well as its value at a time in the past, $t_0$. However, as we show here, conditioning expected entropy on its value at two times rather than one radically changes its dynamics, resulting in a unexpected, very rich structure. 
For example, the expectation value conditioned on two times can have a maximum at an intermediate time between $t_0$ and $t$, i.e., in our past. Moreover, 
it can have a negative rather than positive time derivative at the present. In such ``Boltzmann bridge" situations, the second law would not hold at the present time. We illustrate and investigate these phenomena for a random walk model and an idealized gas model, and conclude with open questions about the models.
\end{abstract}

\maketitle


\section{Introduction}\label{SectionIntroduction}

The seeming paradox of how the second law of thermodynamics can hold despite the
time-symmetry of the laws of physics has been a major subject of debate
for well over a century~\cite{BROWN2009174,WuTy,zermelo1896satz,Loschmidt,davies1974physics}.
Often the paradox is formalized as the question of how the second law of thermodynamics can apparently hold despite (or due to) the time-symmetry of
Boltzmann's H-theorem.
A common claim is that when formulated this way,
the paradox is fully resolved
by the fact that past entropy was lower in the past than it is now (the ``past hypothesis")~\cite{Albert2000,carroll2004spontaneous,wallace2011logic,chen2020}. 

However, this proposed resolution of the paradox is rarely, if ever,  presented in a fully formal manner.
In this paper we correct this oversight.
Since the H-theorem is ultimately a theorem concerning a stochastic process, here we use the theory of Markov processes
to investigate it in a fully formal manner. We prove that the proposed resolution of the paradox in terms of the past hypothesis is wrong as stated, and so is misleading without careful qualifications. 

Going further, our investigation reveals unforeseen, very rich behavior in the stochastic dynamics of (Boltzmann) entropy  through time.
Specifically, consider a statistical system, with entropy $S(t)$ evolving in time. 
Boltzmann's H-theorem can be summarized as the statement that the conditional
expected value of future entropy given current entropy is an increasing function, i.e.,
for all $k > 0$,
\begin{equation}\label{Intro1}
\mathbb{E}\left[S(t + k) \,|\, S(t)\right] > S(t)
\end{equation}
for values of $S(t)$ sufficiently smaller than its maximal value.
By microscopic time-symmetry and time-translation invariance, this immediately
implies that
\begin{equation}\label{Intro2}
\begin{split}
\mathbb{E}\left[S(t + k) \,|\, S(t)\right] &= \mathbb{E}\left[S(t - k) \,|\, S(t)\right]\\
&>0
\end{split}
\end{equation}
for values of $S(t)$ sufficiently smaller than its maximal value, for all $t$.

Let $S(t_0)$ be the entropy at a time $t_0$.  Then Eq. (\ref{Intro1})
means that for values of $S(t_0)$ sufficiently smaller than its maximal value, 
for all $t > t_0$,
\begin{equation}\label{Intro3}
\frac{d}{dt} \mathbb{E}\left[S(t ) \,|\, S(t_0)\right] > 0.
\end{equation}
Importantly,
Eq. (\ref{Intro3}) holds despite the microscopic symmetry of the laws of physics,
encapsulated in Eq. (\ref{Intro2}).

Accordingly, if we are given the fact that in our universe, $S(t_0)$ for a 
time in our past was sufficiently smaller than its maximal value, 
then the time-derivative of the expected entropy of the universe for
all later times, \textit{conditioned
on that previous value of entropy and nothing else}, is non-negative.
This is a fully formal statement of the common 
justification that the second
law follows from the past hypothesis.


However, all axiomatic derivations of the use of probability theory to
reason about uncertainty~\cite{deFinetti1992,Savage1954-SAVTFO-2,Cox1946-COXPFA} say that we should condition all distributions on all that we know, and nothing
that we don't know. This is a fundamental feature of all Bayesian statistics
and Bayesian decision theory.
Furthermore, one might argue that due to the development of modern observational and theoretical cosmology, we know the current value of the universe's entropy
as least as well as its time in some past moment $t_0$.
Combining implies
that when reasoning about the evolution of the entropy of the universe
we should not consider the $t$-dependence of $\mathbb{E}\left[S(t ) \,|\, S(t_0)\right]$
for $t > t_0$,
but rather the $t$-dependence of the ``two-time conditioned'' expectation value,
\begin{equation*}\label{Intro4}
\mathbb{E}\left[S(t ) \,|\, S(t_0), S(t_f)\right]
\end{equation*}
for $t_0 \le t \le t_f$. 

In this paper we investigate the  $t$-dependence of the two-time conditioned 
expected value of the entropy. In Sec. \ref{LemmaSection} we show how to express
the two-time conditioned expected entropy in terms of
two different one-time conditioned expected entropies. This
gives a particularly simple formula for two-time conditioned expected entropy 
whenever the one-time conditioned expected entropy is time-symmetric --- as it is
in the time-symmetric version of Boltzmann's H-theorem in Eq. (\ref{Intro2}).

We then apply this result to analyze the two-time conditioned expected entropy
in toy models of the dynamics of the entropy of the universe. 
The first of these is a random walk model for the evolution
of the state of the universe, and the
second is a gas randomly diffusing between two boxes.

We define ``2nd law violating" curves (2LV) as solutions to $S(t)$ which have a maximum before $t_f$. Intuitively, the 2LV phenomenon is true if $t$ is longer than the relaxation time of the system, or if $S_{now}-S_{past}$ is less that the expected value of the entropy increase (given $S_{past}$ alone) in the $t$ interval. 

These observations suggest that the interpretation of the second law as a simple consequence of a past hypothesis requires, at the very minimum, additional careful qualifications. If we use probability theory properly, conditioning
on what we know and nothing else, then the past hypothesis that $S(t_0)$ was
low does not establish that the second law holds in the 
present. Indeed, when
probability theory is used correctly, a wealth of rich mathematics is
revealed. 

In Sec. \ref{SectionModel1} and Sec. \ref{SectionModel2}, we show that both models have the potential to exhibit changes in the concavity and derivative of $\mathbb{E}\left[S(t ) \,|\, S(t_0), S(t_f)\right]$, in stark contrast to the familiar behavior.  In the first model, this change in concavity common among the space of model parameters. However, in the second model, there is a stronger regularity between 2LV and non-2LV curves that might let us compute the approximate conditions on model parameters that delineate between 2LV and non-2LV behavior. In both models, the two-time conditioned entropy can cross the relaxation curve, which is surprising.

In Sec. \ref{ConcavitySection} and Sec. \ref{delta0Section} we investigate the concavity changes in more detail in addition to showing that setting the final condition to be what is expected from the one-time conditioned distribution alone does affect the two-time conditioned distribution.
Lastly, in Sec. \ref{Conclusion}, we outline open questions about the above models and discuss the relevance of our observations for our understanding of the nature of the second law in our universe. 









\section{From physics to distributions}

We start from a discrete-valued, discrete time Markov process $x$ taking values on a real set $I$ that has a time-symmetric and time-translation invariant one-time conditioned distribution:

\begin{footnotesize}
\begin{equation}\label{DistributionSymmetries}
\begin{array}{cc}
& 
    \begin{array}{cc}
      \mathbf{P}(x(t+k)= a\,|\,x(t)=b) \;=\; \mathbf{P}(x(t-k)= a\,|\,x(t)=b)\\
      \mathbf{P}(x(t+k)= a\,|\,x(t)=b) \;=\;\mathbf{P}(x(t)= a\,|\,x(t-k)=b), \\
    \end{array}
\end{array}
\end{equation}
\end{footnotesize}

\noindent where $a,b \in I$ and $k>0$.

Generally, we want to compute the conditional probability of the stochastic process $x$ conditioned on 2 times: $\mathbf{P}(x(t)\,|\,x(t_0),x(t_f)),$ with $t_0 < t < t_f$.
Critically, we seek time-dependence of expectation values \textit{in between} the initial and final conditions, otherwise we could simply use Markovanity to ignore one of the conditions.

\subsection{Bridge Lemma}\label{LemmaSection}

To evaluate the conditional distribution for the stochastic variable $x$ under two-time conditions, $\mathbf{P}(x(t)\,|\,x(t_0),x(t_f))$, we need to relate it to one-time conditioned distributions like $\mathbf{P}(x(t)\,|\,x(t_0))$, which we assume given.
(Which in our case are provided by Boltzmann's H-theorem and its
extensions.)
We can do this so long as we restrict ourselves to time-symmetric, time-translation
invariant one-time conditioned distributions.



Starting from the two-time conditioned distribution, and using Bayes' theorem, we have:
\begin{equation}\label{Lemma1ProofPart1}
\begin{split}
&\mathbf{P}(x(t)=c\,|\,x(t_0)=a,x(t_f)=b) = \\
&\frac{\mathbf{P}(x(t_0)=a\,|\,x(t)=c,x(t_f)=b)\mathbf{P}(x(t)=c\,|\,x(t_f)=b)}{\mathbf{P}(x(t_0)=a\,|\,x(t_f)=b)},
\end{split}
\end{equation}
with $t_0 \le t \le t_f$.
Markovanity allows us to remove the earlier time conditioning in the two-time conditioned distribution
on the RHS:
\begin{equation}\label{Lemma1ProofPart2}
\begin{split}
&\mathbf{P}(x(t)=c\,|\,x(t_0)=a,x(t_f)=b) = \\
&\frac{\mathbf{P}(x(t_0)=a\,|\,x(t)=c)\mathbf{P}(x(t)=c\,|\,x(t_f)=b)}{\mathbf{P}(x(t_0)=a\,|\,x(t_f)=b)}.
\end{split}
\end{equation}

\noindent To simplify, abbreviate $$I:=(x(t_0)=a), F:=(x(t_f)=b), M:=(x(t)=c),$$ then use Bayes' theorem on each one-time conditioned distribution to write this purely in terms of predictions as:
\begin{equation}\label{Lemma1Simplified}
\mathbf{P}(M\,|\,I,F)=\frac{\mathbf{P}(M\,|\,I)\mathbf{P}(F\,|\,M)}{\mathbf{P}(F\,|\,I)}.
\end{equation}
Note that the denominator is independent of $t$; 
it only depends on the initial and final conditions.

\begin{figure}[h]
\hspace{-0cm}\includegraphics[scale=.5]{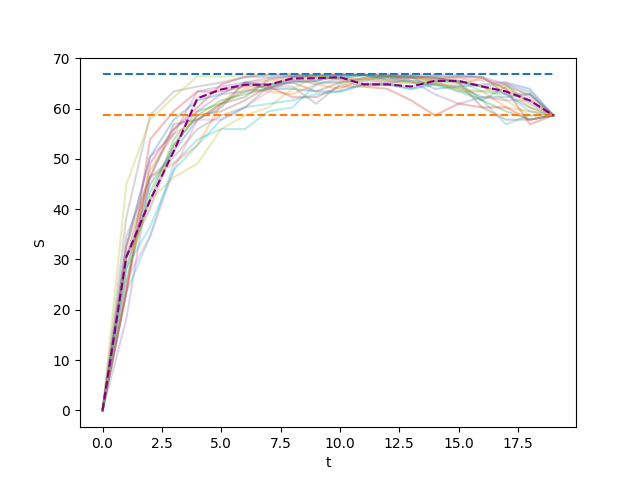}
\caption{An example of a 2LV Boltzmann Bridge in model 1. Many trajectories were sampled with the specified initial and final conditions, and a sample average was taken. Purple dashed: $\mathbf{E}[S(t)\,|\,S(t_0),S(t_f)].$ Blue dashed: $S_{max}.$ Orange dashed: $S(t_f)$.}\label{BridgeSample}
\end{figure}

\subsection{Marginal Distribution}\label{MarginalSection}
Returning to the process $x(t)$ defined in Eq. (\ref{DistributionSymmetries}), what is the marginal $\mathbf{P}(x(t)=a)?$
\noindent  
Using Bayes' theorem and Eq. (\ref{DistributionSymmetries}),
\begin{equation}\label{MarginalProof}
    \frac{\mathbf{P}(x(t+k)=a\,|\,x(t)=b)}{\mathbf{P}(x(t+k)=b\,|\,x(t)=a)}=\frac{\mathbf{P}(x(t)=a)}{\mathbf{P}(x(t+k)=b)}.
\end{equation}
The left hand side of Eq. (\ref{MarginalProof}) is independent of $t$ (under the simple assumption that the one-time conditioned distribution is aperiodic and irreducible),so the right hand side also is, which means the marginal $\mathbf{P}(x(t)=a)$ is time-independent, called stationary.

\section{Modeling}
The symmetries described in Eq. (\ref{DistributionSymmetries}) are in exact analogy to time-reversal and time-translation invariant microscopic physics. To connect with macroscopic physics, we consider a coarse-grained, macroscopic variable $X$ defined in terms of $x$, from which we define the Boltzmann entropy $S:=\log (\Omega(X))$, in parallel to statistical mechanics. $\Omega(X)$ counts the number of microstates $x$ that are compatible with $X$. The corresponding one-time conditioned distributions of $S(t)$ and $X(t)$ will both inherit the same symmetries as Eq. (\ref{DistributionSymmetries}).

For the rest of the paper, we make our distributions for entropy conditioned on values of the macroscopic variable instead of values of the entropy for ease of interpretation and computation, e.g. $\mathbf{E}[S(t)\,|\,X(t_0),X(t_f)]$. This is allowed since the entropy (with a caveat for model 2) is a monotonic function of the macroscopic variable, which allows a unique identification of $S(t)$ and $X(t)$.

In the following discussion, it will be useful to formally define the \textit{relaxation curve} for the entropy:
\begin{equation}
    K(t,X_0):=\mathbf{E}[S(t)\,|\,X(t_0)=X_0]
\end{equation}
and a corresponding relaxation curve for $X$ itself
\begin{equation}
    R(t,X_0):=\mathbf{E}[X(t)\,|\,X(t_0)=X_0].
\end{equation}

\subsection{Model 1}\label{SectionModel1}
Model 1 concerns a random walker that drifts to larger macrostates, first appearing in~\cite{Wolpert2024}.
Consider an unbiased, discrete-time random walker on a $d-$dimensional, uniform, Cartesian lattice with position $\vec{x}$. Let the lattice have a length $2L$, so that the coordinates $x_i$ take integer values in $[-L,L]$. At each timestep, one coordinate among all $x_i$ is chosen at random, then changed by $1,0$ or $-1$ with equal probability. We define a set of coarse-grained macrostates via the `radii' $r:=\text{max}_i \;x_i$, which partition the entirety of the grid. While not a radius in the traditional sense, this quantity labels a set of nested $d-$cubes centered at the origin. The random walker lives on these surfaces and can transit between them. Such a surface has a number of sites $w(r)=(2r+1)^d-(2r-1)^d$, which is a difference of volumes and plays the role of a surface area. We then define a standard Boltzmann entropy via a logarithm of number of microstates, which are the individual sites: $S(r) = \log w(r)$, where $r$ and thus $S$ are time-dependent.

It is a well-known result that the unbiased random walker is ergodic, so the probability of the walker being on any given site is uniform as $t\rightarrow \infty$. 
It is also well-known that unbiased walkers drift to higher distances from the origin on average. That is, $\mathbb{E}[\sqrt{\sum x_i^2}]$ is increasing, even though $\mathbb{E}[\vec{x}]=0.$ Similar is true of the quantity $r$ we have defined, in accordance with the second law.


To make analytic computations, we need the transition rates between the surfaces labeled by $r$, which are computed in the appendix. Note that $r \in [0,L]$.
Formally,
\begin{equation}\label{Model1Rates}
\begin{array}{cc}
& 
    \begin{array}{cc}
      U(r):=\mathbf{P}(r(t+1)=r(t)+1\,|\,r(t))\\
      V(r):=\mathbf{P}(r(t+1)=r(t)-1\,|\,r(t)) \\
      W(r):=\mathbf{P}(r(t+1)=r(t)\,|\,r(t)),
    \end{array}
\end{array}
\end{equation}
with $U+V+W =1$.

\begin{figure}[h]
\hspace{-0cm}\includegraphics[scale=.6]{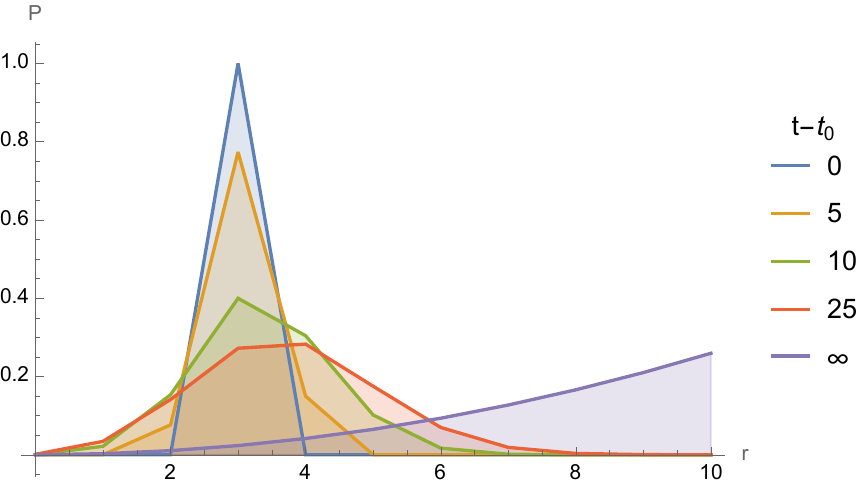}
\caption{For model 1, a plot of the generic time evolution of $\mathbf{P}(r(t)\,|\, r(0)=3)$ with $L=10$, which asymptotes to the stationary distribution Eq. (\ref{Model1StationaryDistribution}).}\label{Model1PMI}
\end{figure}

From the transition rates, Eq. (\ref{Model1Rates}), we have the master equation for model 1, in Einstein notation:
\begin{equation}\label{MasterEquation}
    P_j(t+1)= M_{ij}P_{i}(t),
\end{equation}
where 
   $ M_{ij}:=\mathbf{P}(r(t+1)=i \,|\, r(t)=j)$
is the time-independent $L \times L$ transition matrix, which vanishes if $\,|\,i-j\,|\,>1$. Iterating Eq. (\ref{MasterEquation}) computes the needed one-time conditioned distributions $\mathbf{P}(r(t+k)\,|\,r(t))$, as seen in Fig. \ref{Model1PMI}.

Since probabilities over the sites are uniform as $t\rightarrow \infty$, we can compute the stationary distribution over the coarse-grained cubes as the surface area of the cube divided by the total number of lattice sites: 
\begin{equation}\label{Model1StationaryDistribution}
    \mathbf{P}(r) = \frac{w(r)}{(2L+1)^d}.
\end{equation}
Iterating Eq. (\ref{MasterEquation}) in the infinite limit gets this same result. 
To compute the two-time condition expectation value of the entropy, we take the $\log$ moment of Eq. (\ref{Lemma1Simplified}):
\begin{equation}
\begin{split}
&\mathbf{E}[S(t)\,|\,r(t_0),r(t_f)]=\\
&\sum_{c=0}^L \log w(c)\times\mathbf{P}(r(t)=c\,|\,r(t_0),r(t_f)).
\end{split}
\end{equation}

\begin{figure}
\hspace{.2cm}\includegraphics[scale=.6]{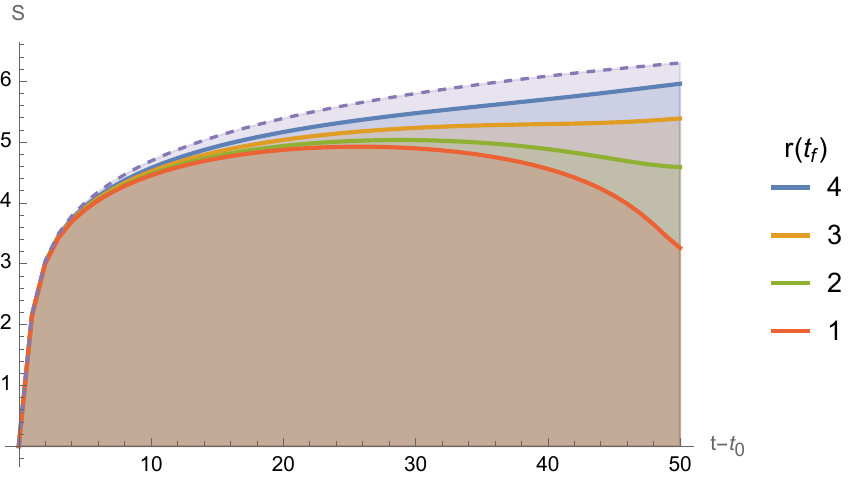}
\caption{For model 1, the time evolution of $\mathbf{E}[S(t)\,|\,r(t_0)=0,r(t_f)=b]$ for various $b$, with $L$=10, $d=3$, and $t_f-t_0=50$. Purple dashed: $K(t,0)$.}\label{Model1Bridges1}
\end{figure}

In most statistical mechanical systems, the equilibrium state is vastly larger than all other macrostates. The second law says that the system's entropy grows to this state's entropy, which is maximum.

\begin{figure}
\hspace{-0cm}\includegraphics[scale=.6]{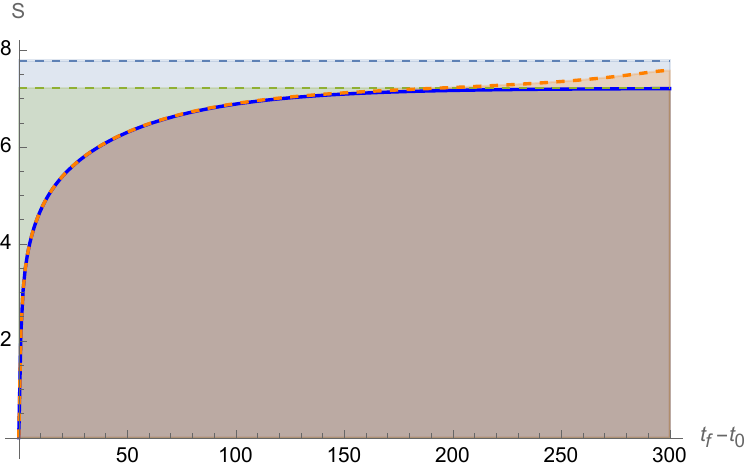}
\caption{For model 1, various S(t) and asymptotes for $L=10$ and $t_f-t_0=300$. Light blue dashed, top: $S_{max}$. Green dashed: $\mathbb{E}[S(t)]$. Orange dashed: $\mathbf{E}[S(t)\,|\,r(t_0)=0,r(t_f)=L-1]$. Dark blue solid: $K(t,0)$.}\label{Model1Relaxation}
\end{figure}

In model 1, the largest macrostate is not necessarily exponentially larger than the other large states. Like any other finite system, the expected value of the entropy of the stationary distribution $\mathbb{E}[S(t)]$ (time-independent), which is what the entropy grows to as $t \rightarrow \infty$, isn't the exactly entropy of the largest state $r=L$, $S_{max}$. 
The two entropies agree as $L$ goes to infinity, however there are interesting features of both two and one-time conditioned entropies that depend on this entropy ``gap" between $\mathbb{E}[S(t)]$ and $S_{max}$ as seen in Fig. \ref{Model1Relaxation}. In the case of the one-time conditioned entropy, the relaxation curve $K(t,a)$ is decreasing if $\log w(a)$ is in the entropy gap - except for an intriguing bump near $t_0$. Surprisingly, $K$ does not need to be monotonic in general, even though $R$ is.

\begin{figure}
\hspace{-0cm}\includegraphics[scale=.6]{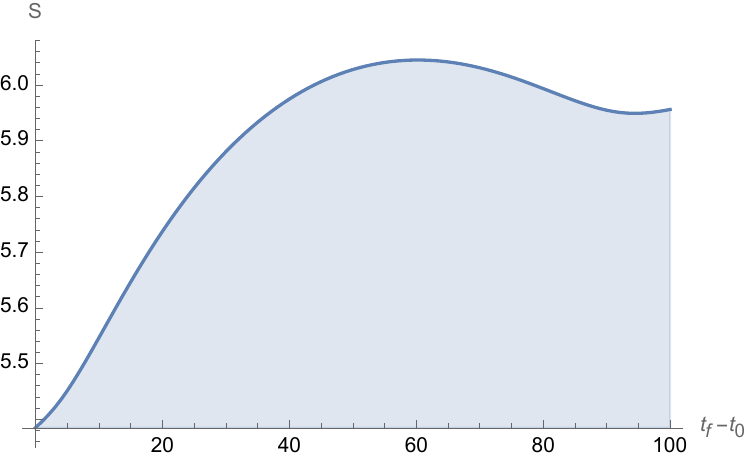}
\caption{For model 1, the time evolution of $\mathbf{E}[S(t)\,|\,r(0)=3,r(t_f)=4]$ with $L$=10, $d=3$, and $t_f-t_0=100$.}\label{Model1Zoomed}
\end{figure}

Fig.~\ref{Model1Bridges1} show examples of $\mathbf{E}[S(t)\,|\,r(0)=0,r(t_f)=b]$ for various values of $b$. It is striking that (upon close inspection) the concavity $\frac{d^2}{dt^2}\mathbf{E}[S(t)\,|\,r(0),r(t_f)]$ can change. There are even extreme cases in which the concavity and/or derivative can change sign \textit{twice}, as seen in Fig. \ref{Model1Zoomed}.



We argued in Sec. \ref{SectionIntroduction} that the 2LV phenomenon can intuitively be seen to be true for systems in which $t_f-t_0$ is much greater than the relaxation time of the system, given that the final condition is below its relaxation value. Two-time conditioned entropies with a final value inside the entropy gap will be concave up near $t_f$ if $t_f-t_0$ is long enough, in complete analogy to how 2LV curves function. Fig. \ref{Model1Relaxation} shows the two-time conditioned entropy ``tracking" the relaxation curve until near $t_f$, when it departs to reach the required final condition (in this scenario and the corresponding 2LV scenario, as $t_f \rightarrow \infty$, the time it takes from the entropy to depart from the relaxation time to achieve the final condition becomes only a function of the final condition.)
However, the concavity changes of the two-time conditioned entropy in Fig. \ref{Model1Bridges1} are for $\log w(r(t_f)) < K(t_f,0)$, so they cannot be understood in terms of the entropy gap, discussed above, but
the concavity change in Fig. \ref{Model1Relaxation} is exactly due to the entropy gap. Both of these phenomena will show up in model 2.

\begin{figure}
\hspace{-0cm}\includegraphics[scale=.6]{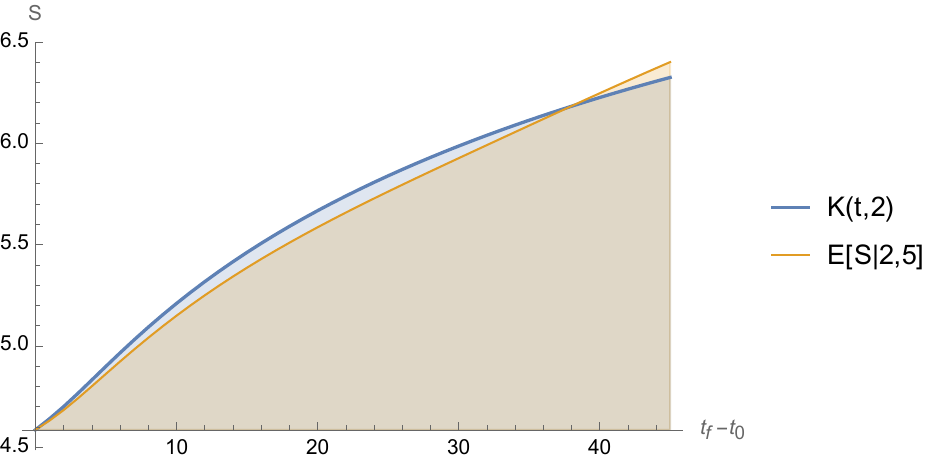}
\caption{For model 1, an example of a two-time conditioned entropy crossing the relaxation curve for the initial condition.}\label{Model1Crossing}
\end{figure}

Among other strange behaviors that model 1 has, the two-time conditioned entropy can cross the relaxation curve $K(t,a)$, as seen in Fig. \ref{Model1Crossing}. Understanding some of these behaviors is non-obvious, and we will return to them in Sec. \ref{ConcavitySection}.

\subsection{Model 2}\label{SectionModel2}
Consider a box of $N$ particles partitioned into left and right, with a small opening in between the two sides that allows particles to switch sides. The system evolves in discrete time, and at each timestep each particle has a probability $\alpha_1$ of switching sides and a probability $\beta_1$ of staying. The microstates are each possible configuration (left or right, for each particle) of distinguishable particles, and the macrostates are defined in terms of the macroscopic variable $n_L$, the number of particles in the left chamber. (We could take $n_R$; it is convention.) As a highly simplified model, we don't consider motion or volume and thus do not compute the thermodynamic entropy, which depends on such quantities. As in model 1, the Boltzmann entropy suffices for our study: 
\begin{equation}\label{Model2Boltzmann}
    S(n_L)=\log \frac{N}{n_L! (N-n_L)!}.
\end{equation} 
Here, $n_L$ is a function of time, thus so is the entropy. There is a subtlety with this entropy that is not present in model 1: a macrostate with $n_L$ particles in the left half and one with $N-n_L$ will have an identical entropy. This is one of the motivating factors of conditioning on $n_L,$ otherwise there would be two two-time conditioned entropy curves satisfying given $S(t_0)$ and $S(t_f)$ - one of the curves passes through $n_L=N/2$ and the other doesn't.
In contrast to model 1, model 2 has an explicit formula for the one-time conditioned distribution, which is derived in the Appendix:
\begin{equation}\label{Model2Conditional}
\begin{split}
&\mathbf{P}(n_L(t+k)=m\,|\,n_L(t)=n) = \\
&\sum_{i=\text{max}(0,n-m)}^{\text{min}(n,N-m)}\binom{n}{i}\alpha_k^i\,\beta_k^{n-i}\times \binom{N-n}{i+m-n}\alpha_k^{i+m-n}\,\beta_k^{N-m-i},
\end{split}
\end{equation}
where $\alpha_k$ is the probability that a given particle is on the opposite side it started from after $k$ steps, and $\beta_k = 1-\alpha_k$ is the probability that a given particle is on the same side as it started after $k$ steps.

The marginal (stationary) distribution is computed via $\lim_{k\to\infty} \alpha_k, \beta_k = 1/2 $ to be 
\begin{equation}\label{Model2Stationary}
        \mathbf{P}(n_L(t)=n)=\binom{N}{n}\frac{1}{2^N},
\end{equation}
which is the binomial distribution with a base probability of $1/2$.


\begin{figure}
\hspace{-0cm}\includegraphics[scale=.6]{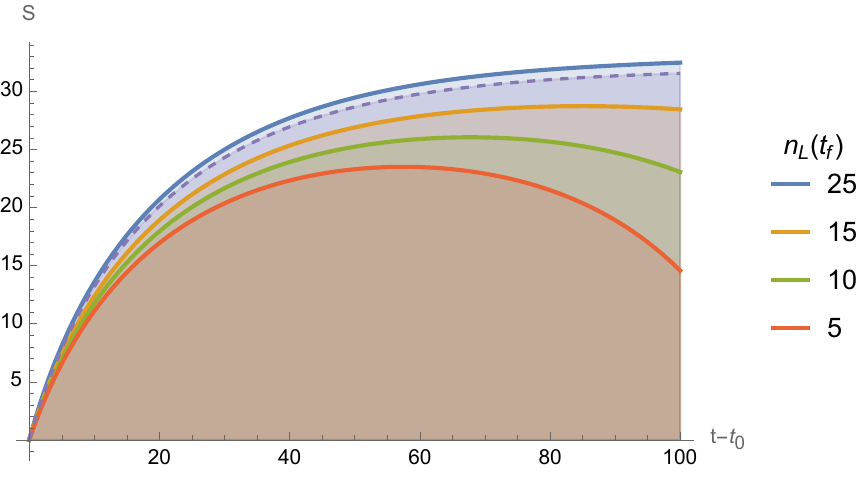}
\caption{For model 2, the time evolution of $\mathbf{E}[S(t)\,|\,n_L(t_0)=0,n_L(t_f)=b]$ for various $b$, with $
\alpha_1=.01$, $N=50$, and $t_f-t_0=100$. Purple dashed: $K(t,0)$. We note the regularity of the curves and that there are no concavity changes.}\label{Model2Bridges1}
\end{figure}


\begin{figure}
\hspace{-0cm}\includegraphics[scale=.5]{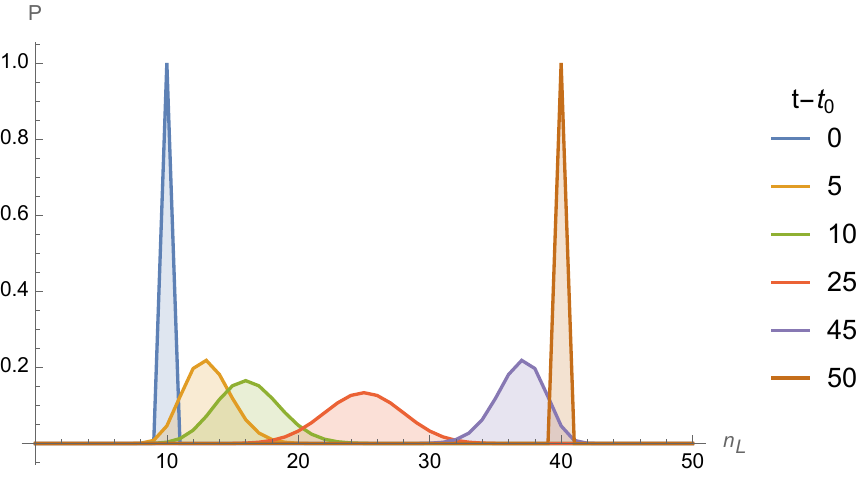}
\caption{For model 2, a plot of the two-time conditioned distribution Eq. (\ref{Lemma1Simplified}) for $n_L(t_0)=10,n_L(t_f)=40,$ and $t_f-t_0 = 50.$ Note that for long $t_f-t_0$, the two-time conditioned distribution approaches the equilibrium distribution for times in between $t_0$ and $t_f$.}\label{Model2Relaxation}
\end{figure}

The relaxation curve grows to its expected value under the stationary distribution, Eq. (\ref{Model2Stationary}), which is the equilibrium entropy Eq. (\ref{Model2Boltzmann}) only in the limit $N\rightarrow \infty$. This is the same effect as discussed in Sec. \ref{SectionModel1}. However, the multiplicity $\binom{N}{N/2}$ of microstates for the largest macrostate grows much more quickly as a function of $N$ than the corresponding quantity for model 1, $w(L)$, does. T
his means that the entropy gap closes more rapidly in model 2 than in model 1 as the systems get larger.

Nonetheless, the concavity $\frac{d^2}{dt^2}\mathbf{E}[S(t)\,|\,n_L(0),n_L(t_f)]$ can change due to the entropy gap.
We also note, unlike model 1, that the relaxation curve $K(t,a)$ is monotonically decreasing if $n_L(t_f) > \mathbb{E}[n_L(t)]$, where the latter is taken with respect to the stationary distribution.

\section{Concavity Changes}\label{ConcavitySection}
In this section, we will prove that the two-time conditioned entropy has a change in concavity under some simple assumptions,. We start with two definitions: define $h:=(S(t_f)-S(t_0))/(t_f-t_0)$, with $S(t_f)>S(t_0)$ and
define the ``Bias" of the process $S(t)$ given an initial and final condition to be $$B(t):=\frac{d}{dt}\mathbb{E}[S(t)\,|\,X(t_0),X(t_f)].$$
Assume $B(t_0)>h$ and $B(t_f)>h$ and let $\{\tau_i\}$ be the set of intersections of $\mathbb{E}[S(t)\,|\,X(t_0),X(t_f)]$ with $h(t-t_0)+S(t_0)$. Finally, let $\tau' = \min{\{\tau_i\}}$ and $\tau'' = \max{\{\tau_i\}}$. Then, there must be a $t_*$ for $t_0 < t_* < \tau'$ such that $B(t_*) < h$, since $\mathbb{E}[S(t)\,|\,X(t_0),X(t_f)]$ must curve down in order to intersect $(t_f, S(t_f))$, and $B(t_0) > h$. 
Define $\tau_{**}$ analogously for $\tau''$ and $t_f$.

We can then conclude that $B(\tau_*) < B(t_0)$. 
So, $dB/dt$ is negative for some contiguous region of $t$ between $t_0$ and $\tau_*$, all assuming $\mathbb{E}[S(t)\,|\,X(t_0),X(t_f)]$ is everywhere twice-differentiable. Similarly we can conclude $B(\tau_{**}) < B(t_f)$, and so $dB/dt$ is positive for some contiguous region of $t$ between $\tau_{**}$ and $t_f$.

As shown in Fig. \ref{Model1Zoomed}, concavity changes can occur without the conditions $B(t_0)>h$ and $B(t_f)>h$ holding, so these conditions should be seen as a sufficient but not necessary.

\section{Does $\delta=0$ provide a correction?}\label{delta0Section}
\begin{figure}
\hspace{-0cm}\includegraphics[scale=.6]{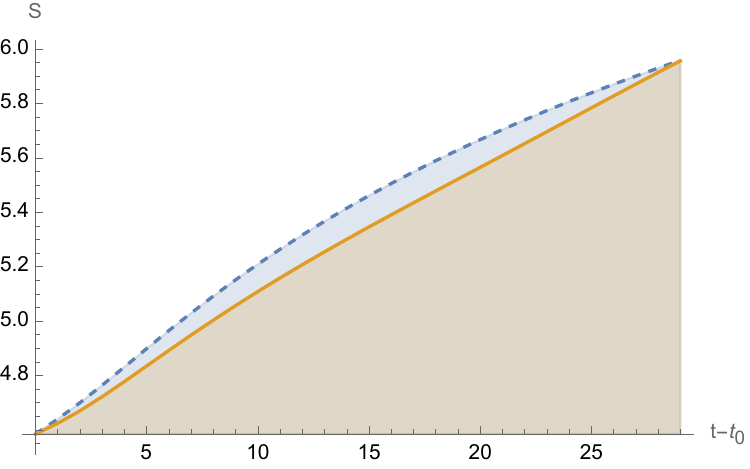}
\caption{For model 1, a comparison of the relaxation curve with the two-time conditioned entropy such that the final condition lies (nearly) exactly on the relaxation curve for $L=10$ and $t_f-t_0=29$. Blue dashed: $K(t,2)$. Orange solid: $\mathbf{E}[S(t)\,|\,r(t_0)=2,r(t_f)=4]$. }\label{Model1Delta0}
\end{figure}

Define $\delta$ as the offset of the final condition from the relaxation curve for $X$: $\delta:=X(t_f)-R(t,X(t_0))$.
It's natural to ask if making the final condition live exactly on its relaxation curve affects two time conditional distribution for the entropy: $$\mathbf{P}(X(t)\,|\,X(t_0),X(t_f)=R(t,X(t_0))=\mathbf{P}(X(t)\,|\,X(t_0)))?$$
Simply evaluating at $t_f$, we find 
\begin{small}
\begin{equation*}
   \mathbf{P}(X(t_f)\,|\,X(t_0),X(t_f)=R(t,X(t_0)))=\begin{cases} 
      1 & X=R(t,X(t_0)) \\
      0 & \text{otherwise} \\
   \end{cases}
\end{equation*}
\end{small}\noindent
whereas $\mathbf{P}(X(t_f)=R(t,X(t_0))\,|\,X(t_0)))< 1$, so it cannot be true. For a more illuminating understanding, we apply Eq. (\ref{Lemma1ProofPart2}) to $\mathbf{P}(X(t)\,|\,X(t_0),X(t_f)=R(t_f,X(t_0)))$ to compute that
the two expressions are related to a time-dependent factor, and trivially these results apply for the distribution for the entropy $\mathbf{P}(S(t)\,|\,X(t_0),X(t_f))$. Though, there is a weaker hypothesis concerning the expectation value:
is it true that
\begin{equation*}
\mathbb{E}[S(t) \,|\, X(t_0), \, X(t_f) = R(t_f,X(t_0))] \neq K(t,X(t_0))? 
\end{equation*}
While there might be an analytic solution, we can answer via numerics that they are not equal, as seen in Fig. \ref{Model1Delta0}, but the effect disappears in the limit $t_f\rightarrow \infty$ as the relaxation curve approaches its asymptote.

\section{Open Questions and Future Work}\label{Conclusion}

Our analysis raises many open question about specific violations and changes of expected 2nd law behavior:
\begin{enumerate}
\item  What is the dependence on $\delta$ of
\begin{equation}\label{OQ1}
\begin{split}
&\max_{t_1 \in [t_0, t_f]} \big(\mathbb{E}[S(t_1) \,\,|\,\, X(t_0), X(t_f) = \delta +\\ &K(t, X(t_0))]
- \mathbb{E}[S(t_1) \,\,|\,\, X(t_0)]\big)?
\end{split}
\end{equation}\noindent
In other words, how does the maximal ``bump upward" due to conditioning on two times rather than one depend on $\delta$?
\item As a variant of Eq. (\ref{OQ1}), what is the dependence on $\delta$ of
\begin{equation}\label{OQ2}
\begin{split}
&\max_{t_1 \in [t_0, t_f]} \dfrac{d}{dt_1} \big(\mathbb{E}[S(t_1) \,\,|\,\, X(t_0), X(t_f) = \delta +\\ 
&K(t, X(t_0))] -\mathbb{E}[S(t_1) \,\,|\,\, X(t_0)]  \big)?
\end{split}
\end{equation}\noindent

In other words, how does the maximal ``weakening of the second law" due to conditioning on two times rather than one depend on $\delta$?

\item What is the maximal $\delta$ such that for some $t_1$ in the allowed interval,
\begin{equation}\label{OQ3}
\frac{d}{dt_1} \mathbb{E}[S(t_1) \,\,|\,\, X(t_0), X(t_f) = \delta + K(t, X(t_0))] < 0
\end{equation}\noindent

i.e., such that the second law is violated?

\item What is the probability of a $\delta$ less than or equal to the maximal one defined in Eq. (\ref{OQ2}) and / or Eq. (\ref{OQ3})?

\item In model 2, we noted the regularity of the Boltzmann Bridges in Fig. \ref{Model2Bridges1}. Define a critical $\delta_c:=X(t_f)-R(t,X(t_0))$ that delineates between 2LV and non-2LV curves. This corresponds to the two-time conditioned entropy being monotonically increasing for $\delta > \delta_c$ and non-monotonic for $\delta < \delta_c$. Then, what is the function $\delta_c(N,
\alpha_1,t_f-t_0,n_L(t_0),n_L(t_f))?$
\item What is the dependence of open questions (1-4) on the model parameters like $t_f - t_0,$ $N$, $\alpha_0$, and $L$?

\item In general the
expected value will not be the median one, so how do all of these results change if rather than define $\delta$ as an offset from the expected value of $r$ or $n_L$,
we define it in terms of percentiles of the cumulative distribution over their values?

\item How do Boltzmann Bridges extend to quantum systems?
\end{enumerate}

Beyond the open questions, we are interested in studying Boltzmann Bridges in cosmology. In particular, although the thermodynamics of matter is non-equilibrium, the maximum entropy of matter in the universe is increasing \cite{Wallace2010-WALGEA}. Much more in-depth analysis needs to be done to have more realistic modeling of the universe - see \cite{Egan_2010} for a study of theoretical predictions of the universe's entropy over time. Though such analysis might be possible, almost surely in the limit of a large number of particles and a large number of degrees of freedom, the set of values of $\delta$ under which the second law holds has probability infinitesimally close to 1. Despite this, we have raised a careful correction to the standard formulations of the 2nd law from using past hypothesis. The correction can be quite large for the smaller systems we study in this paper, which might have relevance to real microscopic and mesoscopic systems. 


\section{Appendix}
\subsection{Model 1 Transition Rates}
To compute the probabilities in Eq. (\ref{Model1Rates}), we first need to compute the same probabilities conditioned on $\vec{x}$, the location of the random walker. For $0<r<L$, define
\begin{equation}
\begin{array}{cc}
& 
    \begin{array}{cc}
      u(\vec{x}):=\mathbf{P}(r(t+1)=r(t)+1\,|\,\vec{x})\\
      v(\vec{x}):=\mathbf{P}(r(t+1)=r(t)-1\,|\,\vec{x}) \\
      w(\vec{x}):=\mathbf{P}(r(t+1)=r(t)\,|\,\vec{x}).
    \end{array}
\end{array}
\end{equation}

Let $\eta$ denote the number of maximum elements among the $x_i$. The probability to choose one of them is $\eta/d,$ and the probability of that choice increasing is $1/3$, so
\begin{equation}\label{u}
    u(\vec{x})= \frac{1}{3}\frac{\eta}{d}.
\end{equation}

Computing $v$ is similar: one of the maxima must be chosen, and that must decrease. However, $r$ only decreases if the maxima among $x_i$ is unique, so:

\begin{equation}\label{v}
v(\vec{x}) =
    \begin{cases}
        \frac{1}{3d} & \text{if } \eta = 1\\
        0 & \text{otherwise }
    \end{cases}
\end{equation}

Thus $v$ is $0$ unless there is only one maximum, and it equals $u$ otherwise. Thus, $u \geq v$. Now, we must compute the expectation value of $u,v$ over all sites on a given $r$ in order to compute $U(r), V(r),$ for $0<r<L$.

Given $r$, the probability of any given site is uniform over the surface of the $d$-cube, which is $\mathbf{P}(\vec{x}\,|\,r(\vec{x}))=1/w(r).$ Now, we compute how many sites on a given $r$ surface have $g$ maxima. This is critical since if a site has more than $1$ maximum, it lives on a hyper-edge and the radius cannot decrease.

A $d$-cube has a multiplicity $E_{g,d}$ of $g$ dimensional faces, where $E_{g,d}=2^b\binom{d}{g}$. The number of sites living on one of these $b-$faces comes from the multiplicity times the volume (not overcounting) of that face, $$w_g(r)=2^g \binom{d}{g}(2r-1)^{d-g}.$$
We can prove via the binomial theorem that these faces partition the surface of constant $r$:
$$\sum_{g=1}^d w_g(r)=w(r).$$

\begin{figure}
\hspace{-0cm}\includegraphics[scale=.6]{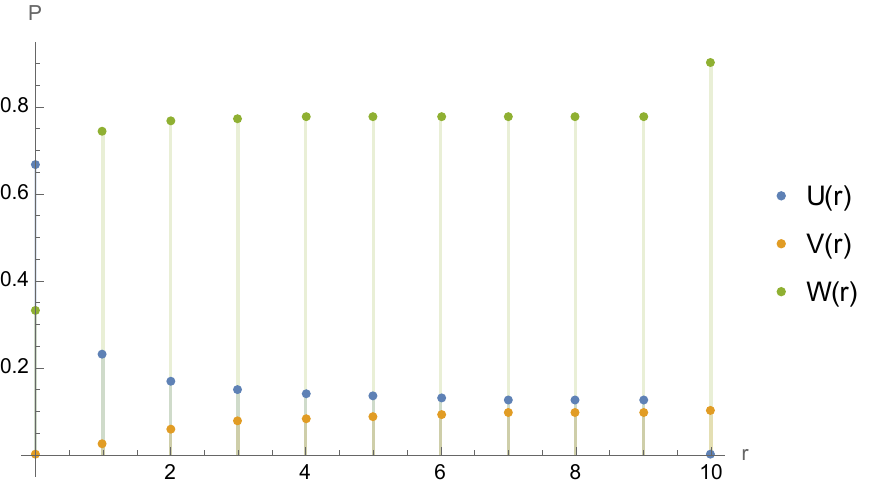}
\caption{For model 1, plot of the $r$ dependence of $U(r)$, $V(r)$, $W(r)$ for $L=10$. Boundary conditions ensure that $U(L)=0$. It is easy to compute that $U,V\rightarrow 1/(3d)$ as $L\rightarrow \infty$.}
\label{Model1RatePlot}
\end{figure}

A $g-$face is $d-g$ dimensional. So $g=0$ corresponds to the entire volume; $g=1$ corresponds to faces (which touch no edge); $g=d-1$ corresponds to $1$-dim edges, and $g=d$ corresponds to 0-dim corners. We can compute $V$ with Eq. \ref{v} and replacing $(\text{if } \eta =1)$ with $\mathbf{P}(\eta = 1)$, where the probability is uniform over the lattice among all allowed sites. Then, 
\begin{equation}\label{BigV}
    V(r)=\frac{1}{3d}\frac{w_1(r)}{w(r)},
\end{equation}
where $V(r)$ is the averaged quantity, the probability of the radius decreasing during a timestep. $U(r)$ is computed similarly:
$U(r)=\frac{1}{w(r)}\times\sum_{\vec{x}}u(\vec{x}).$ Then,
\begin{equation}\label{BigU}
U(r)=\frac{1}{3d}\frac{1}{w(r)}\sum_{g=1}^d w_g(r)\times g.
\end{equation}
It is simple to show that $U(r) \geq V(r).$ On the boundaries $r=0, r=L$, we have the probabilities $$U(0)=2/3,\, V(0)=0,\, U(L)=0,$$ independent of $d$.
We have chosen the boundary condition that there is a maximum $L$. We make the simple choice that if a random walker were to increase in radius from $L$, instead it doesn't move. Then, $V(L)$ is what is expected from Eq. (\ref{BigV}), and $W(L)$ changes so that $V(L)+W(L)=1$. This is in contrast to reweighting $v(L),w(L)$ based upon the convention that the random walker randomly selects uniformly over adjacent sites, which would result in $V(L)$ being greater. An example of the transition rates can be seen in Fig. \ref{Model1RatePlot}.

\subsection{Model 2 Conditional Distributions}
We start from the Markov process $s(t)$, taking values $L$ or $R$, which denotes the position of a single particle within the partitioned box. The process $s(t)$ has the symmetries of Eq. (\ref{DistributionSymmetries}), and all following equations are symmetric under interchanging $L \leftrightarrow R$. 

We start with the one-time conditioned probabilities:
\begin{equation}\label{Model2onestepconditionals}
\begin{array}{cc}
& 
    \begin{array}{cc}
      \mathbf{P}(s(t+1)=L\,|\,s(t)=R)=\alpha_1\\
      \;\mathbf{P}(s(t+1)=L\,|\,s(t)=L)=\beta_1.\\
    \end{array}
\end{array}
\end{equation}
We want to compute arbitrary one-time conditioned probabilities of the form 
$ \mathbf{P}(s(t+k)=L\,|\,s(t)=R),$
where $k>0$. As shorthand we define
\begin{equation}\label{Model2onetimeconditionals}
\begin{array}{cc}
& 
    \begin{array}{cc}
      \mathbf{P}(s(t+k)=L\,|\,s(t)=R)=\alpha_k\\
      \; \mathbf{P}(s(t+k)=L\,|\,s(t)=L)=\beta_k.\\
    \end{array}
\end{array}
\end{equation}
Markovian dyanmics
gives the following recursion relations among these quantities:
\begin{equation}\label{Model2RecursionRelations}
\begin{array}{cc}
& 
    \begin{array}{cc}
      \alpha_{k+1}=\beta_1 \alpha_k + \alpha_1 \beta_k\\
      \beta_{k+1}= \alpha_1 \alpha_k + \beta_1 \beta_k.
    \end{array}
\end{array}
\end{equation}
The first follows from expanding the probability that a particle is on
the left at timestep $t + k + 1$ in terms of the probabilities
that in that timestep it didn't switch sides but was also on
the left at the end of the preceding timestep, plus the probability that
it did switch but was on the right side at the preceding timestep.
The second equation follows by using analogous reasoning.
Eq. (\ref{Model2RecursionRelations}) has solutions
\begin{equation}\label{RecursionRelationSolutions}
\begin{array}{cc}
& 
    \begin{array}{cc}
      \alpha_{k}=\frac{1}{2}(1-(\beta_1-\alpha_1)^k)\\
      \;\beta_{k}= \frac{1}{2}(1+(\beta_1-\alpha_1)^k).
    \end{array}
\end{array}
\end{equation}

We next use Bayes' theorem to compute retrodictive one-time conditioned distributions of the form 
$\mathbf{P}(s(t-k)=L\,|\,s(t)=R)$, where $k>0$ again. Via time-translation symmetry, this is $\mathbf{P}(s(t-k)=L\,|\,s(t)=R) =\mathbf{P}(s(t)=L\,|\,s(t+k)=R)$. Via Bayes' theorem on the right hand side, we have
\begin{equation}\label{Model2DerivationBayes}
\begin{split}
&\mathbf{P}(s(t)=L\,|\,s(t+k)=R) = \\
&\frac{\mathbf{P}(s(t+k)=R\,|\,s(t)=L) \mathbf{P}(s(t)=L)}{\mathbf{P}(s(t+k)=R)}.
\end{split}
\end{equation}
Due only to Eq. (\ref{DistributionSymmetries}), which are the symmetries of Eq. (\ref{Model2onestepconditionals}), the marginals $\mathbf{P}(s(t)=L)$ and $\mathbf{P}(s(t+k)=R)$ have to be time-independent, which we proved in Sec. \ref{MarginalSection}. The only way for this to be possible while conserving probability is $\mathbf{P}(s(t)=L)=\mathbf{P}(s(t)=R)= 1/2$, as expected.

We now compute the general one-time conditioned distribution, Eq. (\ref{Model2Conditional}). Denote the number of left to right transitions over $k$ timesteps to be $T^{LR}_k.$ First, we compute the probability that $T^{LR}_k=p$, given that $n_L(t)=n.$ Among the $n$ particles, choosing $p$ to switch gives a factor of $\binom{n}{p}$. The probability that any given of those particles is on the other side after the $k$ steps is $\alpha_k$. The probability that any of the rest of the $n-p$ particles are on their starting side is $\beta_k$. Multiplying the constituent probabilities and exponentiating to account for each particle, we have 
\begin{equation}\label{Model2derivation2}
\mathbf{P}(T^{LR}_k=p \,|\, n_L(t)=n)=\binom{n}{p}\;\alpha_k^{p}\;\beta_k^{n-p}.
\end{equation}

Now, if we compute the probability that there are $p$ left to right and $q$ right to left transitions over $k$ steps, we can simply sum over that probability to compute Eq. (\ref{Model2Conditional}).
The probability is a product of two applications of Eq. (\ref{Model2derivation2}):
\begin{equation*}\label{Model2Derivation3}
\begin{split}
&\mathbf{P}(T^{LR}_k=p, T^{RL}_k=q )=\\
&\binom{n}{p}\;\alpha_k^{p}\;\beta_k^{n-p} \times \binom{N-n}{q}\alpha_k^{q}\beta_k^{N-n-q}.
\end{split}
\end{equation*}
Enforcing conservation of particle number $m-n=q-p$ and the inequalities $0 \leq q \leq N-n$ and $0 \leq p \leq n$ gets us the desired conditional distribution Eq. (\ref{Model2Conditional}).

\begin{acknowledgements}
    The authors would like to express thanks to Wayne Myrvold for guidance on analytical computations for model 2. The authors would also like to thank Anthony Aguirre for helpful discussions. This work was supported by funding from the Santa Fe Institute.
\end{acknowledgements}

\bibliography{carlo-version}

\end{document}